\documentclass[11pt]{article}

\usepackage{fullpage}
\usepackage{epic,eepic,epsfig,graphics,amsmath,color,fullpage}
\usepackage{latexsym}
\usepackage{algorithm}
\usepackage{algorithmic}
\usepackage{makecell}
\usepackage{tabularx}
\usepackage[hang]{subfigure}
\usepackage{times}
\usepackage{graphicx}

{\renewcommand{\baselinestretch}{1}

\begin{document}


\newcommand{\cardi}[1]{ \vert #1 \vert}

\def\vval{{{\#}_{v}}}
\def\estval{{{\#}_{est_i}}}
\def\botval{{{\#}_{\bot}}}

\newtheorem{definition}{Definition}
\newtheorem{theorem}{Theorem}
\newtheorem{notation}{Notation}
\newtheorem{lemma}{Lemma}
\newtheorem{corollary}{Corollary}
\newcommand{\toto}{xxx}
\newenvironment{proofT}{\noindent{\bf
Proof }} {\hspace*{\fill}$\Box_{Theorem~\ref{\toto}}$\par\vspace{3mm}}
\newenvironment{proofL}{\noindent{\bf
Proof }} {\hspace*{\fill}$\Box_{Lemma~\ref{\toto}}$\par\vspace{3mm}}
\newenvironment{proofC}{\noindent{\bf
Proof }} {\hspace*{\fill}$\Box_{Corollary~\ref{\toto}}$\par\vspace{3mm}}

\newenvironment{theorem-repeat}[1]{\begin{trivlist}
\item[\hspace{\labelsep}{\bf\noindent Theorem~\ref{#1} }]}%
{\end{trivlist}}

\newenvironment{lemma-repeat}[1]{\begin{trivlist}
\item[\hspace{\labelsep}{\bf\noindent Lemma~\ref{#1} }]}%
{\end{trivlist}}

\newcounter{linecounter}
\newcommand{\linenumbering}{(\arabic{linecounter})}
\renewcommand{\line}[1]{\refstepcounter{linecounter}
\label{#1}
\linenumbering}
\newcommand{\resetline}{\setcounter{linecounter}{0}}

\newcommand{\grumbler}[2]{\begin{quote}{\sl \bf #1:} #2\end{quote}}


\title{\bf Self-stabilizing Algorithm for Maximal Distance-2 Independent Set}

\author{Badreddine Benreguia$^{\ddag}$~~
        Hamouma Moumen$^{\ddag}$\footnote{Address for correspondence: H. Moumen, Department of Informatics, University of Batna 2, Fesdis 05078, Batna, Algeria.}~~
        Soheila Bouam$^{\ddag}$~~
        Chafik Arar$^{\ddag}$\\~\\
$^{\ddag}$  University of Batna 2, 05000 Batna, Algeria \\
{\small {\tt h.moumen@univ-batna2.dz  }} }

\date{}
\maketitle
\renewcommand{\baselinestretch}{1,0}

\begin{abstract}

In graph theory, an independent set is a subset of nodes where there are no two adjacent nodes. The independent set is maximal if no node outside the independent set can join it. In network applications, maximal independent sets can be used as cluster heads in ad hoc and wireless sensor networks. In order to deal with any failure in networks, self-stabilizing algorithms have been proposed in the literature to calculate the maximal independent set under different hypothesis.In this paper, we propose a self-stabilizing algorithm to compute a maximal independent set where nodes of the independent set are far from each other at least with distance 3. We prove the correctness and the convergence of the proposed algorithm. Simulation tests show the ability of our algorithm to find a reduced number of nodes in large scale networks which allows strong control of networks.
~
\end{abstract}







\section{Introduction}
\label{intro}
 \subsection{Context of the study and motivation}
 \label{context}
Self-stabilization is a fault tolerance approach that allows distributed systems to achieve a global correct configuration starting from an unknown initial configuration. Without external intervention, a self-stabilizing algorithm is able to correct the global configuration of the distributed system in a finite time. Various self-stabilizing distributed algorithms have been proposed in the literature using graph theory such as leader election, nodes coloring, domination problem, identifying the independent set, constructing the spanning tree. These algorithms have many benefits in the real-life applications, for example independent sets have been used as cluster heads in ad hoc and sensor networks \cite{AAD19, AD19, BDJV05,LT11}.

In graphs, independence is commonly defined as follow: let $G=(V,E)$ be a graph, where $V$ is the set of nodes and $E$ is the set of edges. An independent set is a subset of nodes $S \subset V$ such that there is no two adjacent nodes in $S$. The distance between any two nodes in $S$ is greater than 1. An independent set $S$ is said $maximal$, if there is no superset of $S$ that could be an independent set.  In other words, there is no node outside the maximal independent set (MIS) that may join MIS.
It is well known in graph literature that MIS is considered also as dominating set because every node out of MIS has at least a neighbor in MIS (every node outside MIS is dominated by a node of MIS).

In this paper, we deal with a particular case of independent set. We call $S$ maximal distance-2 independent set (MD2IS), if of $S$ nodes are independent and the distance between any two nodes among them is strictly greater than 2.
Figure 1 illustrates difference between MIS and MD2IS where green nodes are independent. Observe that in MIS (a), distance of 2 could be found between independent nodes. However, the distance between green nodes in MD2IS (b) is strictly greater than 2.

\begin{figure}[hbt!]
\begin{center}
\includegraphics[width=4in]{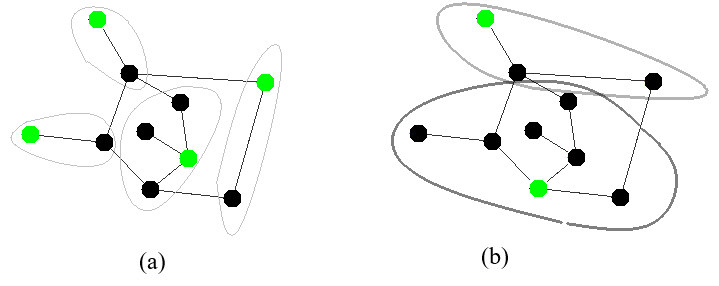}
\caption{(a) Maximal independent set. (b) Maximal distance-2 independent set.}
\end{center}
\label{example_MIS_MD2IS}
\end{figure}

Nodes of MIS are used as servers (cluster heads) in ad hoc and wireless sensor networks to provide important services for other nodes. Each cluster head has to guarantee services for nodes connected to it, that are called members of the cluster. Cluster members represent nodes outside of MIS. A cluster head could serve its members by routing information, providing keys of encryption, giving names for members,...

Figure 1(b) shows that elements of MD2SI could be used as cluster heads where members connected to the head could be within distance of 2. However, using MIS, members could not be located within distance more than 1. Obviously, MD2IS provides a more reduced number of clusters than MIS.
The choice of the cluster heads is important in order to contribute in extending lifetime of wireless sensor and ad hoc networks. Using MD2IS rather than MIS as cluster heads could provide a good alternative in this sense especially that lifetime is the major problem of these networks. In addition to that and in order to deal with any possible failure, we use self-stabilizing algorithm that ensures reconstructing cluster heads, after the failure occurs, allowing the network still operational.

Finding the maximal independent set (MIS) in graphs using self-stabilization paradigm was studied in literature for the first time by Shukla et al. in 1995 \cite{SRR95}. Authors have used a straightforward idea based on two rules: (1) a node $v$ joins the set $S$ (which is under construction) if $v$ has no neighbor in $S$, and (2) a node $v$ leaves the set $S$ if at least one of its neighbors is in $S$.
Other variants of self-stabilizing algorithms constructing independent set have been introduced to deal with particular problems which try to minimize the algorithm complexity \cite{Tur07} or to be suitable for distributed daemon \footnote{See section \ref{daemon notion}.} \cite{GHJS03,IKK02}. Reader can refer to the survey \cite{GK10} for more details on MIS self-stabilizing algorithms.
Other self-stabilizing algorithms have been proposed for independent sets imposing additional constraints besides to the independence condition. For example, \cite{NGHK15} has presented an algorithm to discover the independent set where each node $u$ out of $S$ has at least a neighbor $v$ in $S$ such that $deg(v)>deg(u)$. In \cite{AD19} authors propose a distributed self-stabilizing algorithm to find MIS using two-hop (distance-2) information in wireless sensor networks.

\subsection{Related works}
\cite{Joh14,Joh15} has proposed a self-stabilizing algorithm to find the independent dominating set imposing a distance greater than $k$ between any two nodes of the independent set. Work \cite{Joh15} is an improvement of the memory management regarding the first one \cite{Joh14}. Every node outside the independent set is within distance $k$. \cite{DDL19} presented a self-stabilizing algorithm to compute a dominating set $S$ (which is not independent) where every node out of $S$ has to be distant from $S$ at most by $k$. Although the precedent algorithms have bounded complexity $O(n+k)$ in rounds, authors indicate that these algorithms might still never converge under the distributed daemon, since the daemon could ignore an enabled nodes. It is known in literature that: if the round complexity of a self-stabilizing algorithm is finite, this does not mean it converges \cite{DDL19}. Therefore, the computation of the convergence time still an open question \cite{DDL19,Joh14,Joh15} for independent (or dominating) set at distance $k \geq 2$.

\subsection{Contribution}
In this paper, we propose a self-stabilizing algorithm to find maximal distance-2 independent set called MD2IS. We prove the correctness and the convergence of the presented algorithm. Using a serial central daemon, and starting from an arbitrary configuration, MD2IS reaches the correct configuration in a limit number of moves. The serial central daemon allows reaching the correct configuration in the worst case at $2n$ moves. Which means that MD2IS converges in $O(n)$ moves using a central daemon under expression distance-2 model. For distance-one model, our algorithm reaches the correct configuration in $O(nm)$ moves using distributed daemon, where $n$ is the nodes number and $m$ is the edges number. Proofs and simulation tests confirm the convergence of the proposed algorithm that provides smaller independent sets in large scale graphs.

Having independent set more reduced is useful in many applications and allows more control on large scale networks.
For example, the problem of locating an anonymous source of diffusion \cite{PTV12,SMA12} needs the placement of few nodes as observers \cite{SCT16}. A reduced number of nodes that occupy important locations in graphs, can be used as observers to detect the source of rumors in social networks.

\subsection{Organization of the paper}
This paper is made up of five sections. Section 2 presents the model and the terminology used for self-stabilization. In section 3, we introduce the proposed self-stabilizing algorithm for finding a maximal independent set at distance-2. Simulation tests are conducted in section 4. Finally, section 5 concludes the paper.

\section{Model and terminology}
Networks and distributed systems are modelled generally as an undirected graph $G =(V, E)$ where the units of processing represent the set of nodes $V$ and the links are the set of edges $E$. The neighborhood of a node $v \in V$ is defined as $N(v)=\{u \in V : vu \in E\}$. Usually, we say that two nodes $v,u$ are $adjacent$ if $u \in N(v)$. We define the neighborhood at distance-2 as $N(v)_{dist2}=N(v) \cup \{e \in V : (\exists u \in N(v) : e \in N(u) )\}$. It is known in the graph literature that the distance between two adjacent nodes is 1. Clearly, the $v$'s neighborhood at distance-2 gathers the set of nodes within distance of 1 and 2.

A set $S \subset V$ is $independent$ if no two nodes of $S$ are adjacent. In other word, there is no two nodes in $S$ at distance 1. Generally, a set $S$ of nodes is $distance$-$k$ independent if every node in $S$ is distant at least $k+1$ to any other node of $S$ \cite{Joh14}. Consequently, a distance-2 independent set is a subset $S$ of $V$ where every two nodes of $S$ are at distance $>2$. Recall that in the usual case of MIS, each node in the graph is either independent or dominated $i.e.$ every node of MIS is independent and every node outside MIS is dominated \cite{HHJS03}. 
In our case, every node $u$ out of MD2IS is either dominated by a node $v \in $ MD2IS where $dist(u,v)=1$ or is dominated by $v$ where $dist(u,v)=2$. 
Note that, a node $u$ out of MD2IS could be dominated by many nodes of MD2IS, for example it is possible to find a node $u$ dominated by $v1$ and $v2$ where $dist(u,v1)=1$ and $dist(u,v2)=2$.

\textbf{Definition} : A $distance$-$2$ $independent$ $set$ of a graph $G(V,E)$ is a subset $S$ of $V$ such that the distance between any two nodes of $S$ is strictly greater than $2$. $S$ is $maximal$ if no superset of $S$ is also a distance-2 independent set.

An algorithm is self-stabilizing if it can (1) reach a global correct configuration called $legitimate$ (2) during a finite time after it starts from an unknown configuration. When a self-stabilizing algorithm reaches the correct configuration it must stay inside the correct configuration (known as $closure$ condition). Hence, to show that an algorithm is self-stabilizing, it is sufficient to prove its $closure$  for the legitimate configuration and its $convergence$ to achieve the desired configuration in a finite time.
In the \textit{uniform} self-stabilizing system, all nodes execute the same code which is a set of rules having the form: $\textbf{if}$ $guard$ $\textbf{then}$ $statement$ (written as: $guard \longrightarrow statement$). In this case, nodes use the same local variables that describe their $state$.  The \textit{guard} is a (or a collection of) boolean expression. If a guard is evaluated to be true, the corresponding rule is said $enabled$ (or $priviliged$). We say that a node is $enabled$, if at least one of its rules is enabled.

Executing the statement of the enabled rule by the node is called a $move$. An enabled node can make a move only if it is selected by a scheduler called a $daemon$. A move allows updating the local state (local variables) in order to make the node more legitimate with its neighborhood.

\subsection{Daemon notion}
\label{daemon notion}
The execution of self-stabilizing algorithms is managed by a daemon (scheduler) that selects the enabled nodes to move from a configuration to another configuration. Two types of daemons are widely used in literature: central and distributed daemons. In the central daemons, only one enabled node is selected to be moved among all the enabled nodes. The central daemon, called also serial, allows enabled nodes executing a move (one by one) by a serial order. However, in the distributed daemons, a subset of nodes are selected among the set of privileged nodes to make a move simultaneously. The selected subset of nodes to be moved simultaneously forms a $round$. The distributed daemon is said $synchronous$ when all the enabled nodes are selected to move simultaneously.

\subsection{Distance model}
Generally, most of the existing self-stabilizing algorithms, use the distance-one model wherein each node has a partial view on neighbors at distance one (through an access to the $state$ variable of the neighbors). However, there is other distance-2 models (like expression model) where every node has to get information of its neighborhood at distance-2. In the expression model which is particular case of distance-2 model, access to distance-2 is reached indirectly through access to $expression$ of neighbors. In the distance-2 model, hypothesis should be constructed carefully. For example, to the best of our knowledge, there is no algorithm of distance-2 that can operate under distributed daemon. The existing algorithms of distance-2 have been developed only under central daemon.

\subsection{Transformers}
A common approach, known in literature \cite{Tur12,GS12}, allows converting a self-stabilizing algorithm $A$ operating under a given hypothesis to a new algorithm $A^{T}$, that operates under other hypothesis different from the first ones. However, the transformation guarantees that the two algorithms converge to the same legitimate configuration. Different kinds of transformers can be found in literature like \textit{distance} transformers and \textit{daemon} transformers. Generally, an overhead of the algorithm complexity is caused by the transformation which leads generally to a slowdown in the convergence of the transformed algorithms.
In this paper, we use the transformer proposed by \cite{Tur12} that allows to transform any self-stabilizing algorithm operating under serial central daemon and expression distance-2 model to a self-stabilizing algorithm that operates under distributed daemon and distance-one model.

\subsection{Our execution model}
In this paper, we develop a uniform self-stabilizing algorithm. In a first step, we suppose our algorithm operates under expression distance-2 model using central daemon. After that, we use transformer proposed by \cite{Tur12} that allows converting our algorithm to another algorithm that runs under distance-one model using distributed daemon.

\section{Self-stabilizing algorithm MD2IS}
The proposed self-stabilizing algorithm presented in Algorithm \ref{alg1}, allows finding the maximal distance-2 independent set.
Each node $v$ maintains a local variable $state$ and an expression $exp$. The $state$ variable could take one of the values $In$ or $Out$.
Once the system reaches the legitimate configuration, all the nodes are disabled and the set $S=\{v \in V : v.state = In \}$ forms the maximal distance-2 independent set.
However, in the illegitimate configuration, a serial moves of enabled nodes are executed until the global correct configuration is reached.
 At every move, an enabled node is selected randomly by the central daemon.

Every node checks its state regarding its neighborhood using
$state$ and $exp$. Expression $exp$ is used to calculate the
number of the neighbors in $S$. Generally, the expression model
allows discovering the neighborhood at distance-2. Once a node
could read expressions of its neighbors, it will have an
information about its neighbors at distance-2.

In our algorithm, R1 shows that if a node $v$ out of $S$ reads the $state$ and the $exp$ of all its neighbors and finds that all the $state=Out$ and all the $exp=0$, hence all the neighbors at distance-2 are out, therefore node $v$ has to join the set $S$. Conversely, R2 illustrates that if a node $v \in S$ knows that there exists at least one of its neighbors $u$ such that $u.state=In$ or $u.exp>1$, thus $v$ has to leave $S$. Note that if $v \in S$ and $u.exp>1$ means that $u$ is dominated at least by two nodes: $v$ and another node. Observe also that if $v.state=IN$ and $u.exp=1$, this means that $u$ is dominated only by $v$ and it is impossible that $u.exp=0$ because there exists at least $v \in S$ as a neighbor of $u$.

It is clear that R1 ensures the independence property because node $v$ will be independent at distance-2 by executing R1 leading to $v.state=In \wedge \forall u \in  N(v) : (u.state=Out \wedge u.exp=0)$. However, R2 guarantees that every node $v$ out of MD2IS is dominated at least by distance-2 because ($v.state=Out \wedge \exists u \in  N(v) : (u.state=In \vee u.exp>1)$).

Figure 2 shows how our algorithm converges from an initial illegitimate configuration to a final legitimate configuration using a central serial daemon. Nodes outlined in red are (privileged) candidates to execute a move. The central daemon selects in every step only one privileged node to execute a move. Green nodes represents MD2IS which is reached after 5 moves by the sequence of rules: R2, R2, R2, R1, R1. Observe that in some cases, a move (for example transition from configuration c to configuration d) can make other nodes enabled in the next configuration.

\begin{figure}[hbt!]
\begin{center}
\includegraphics[width=6in]{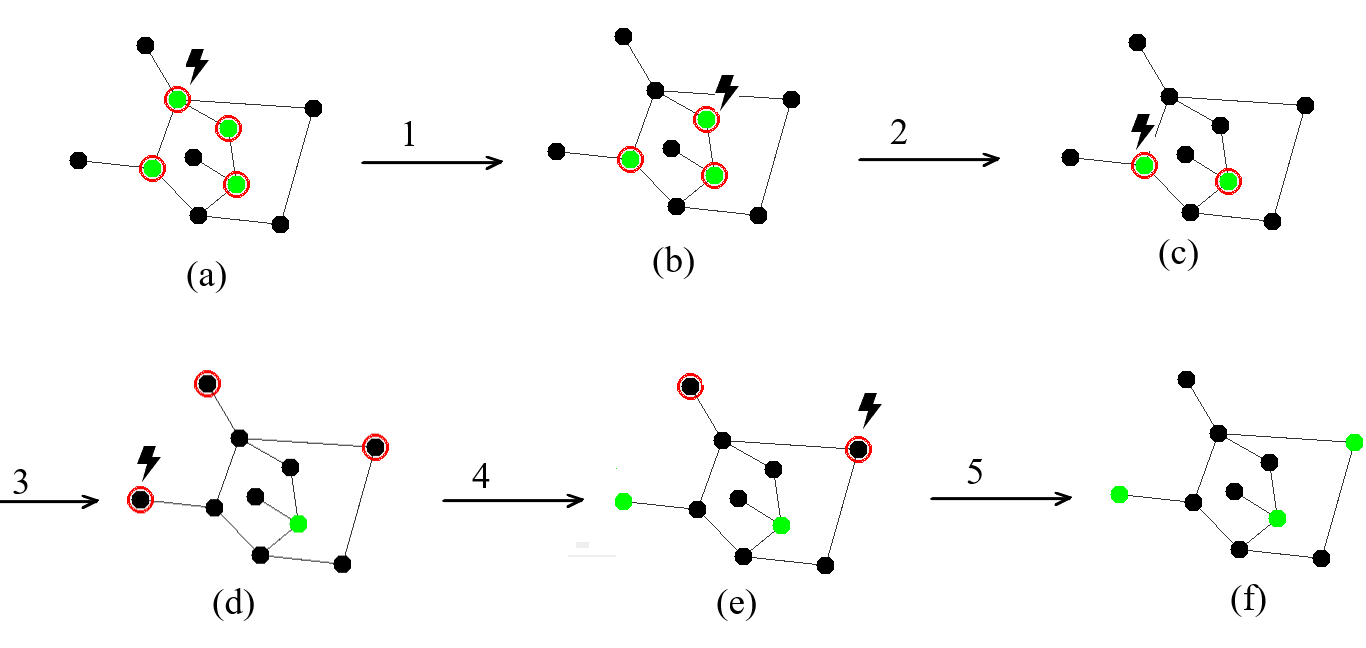}
\caption{Convergence to the final legitimate MD2IS configuration.}
\end{center}
\label{fig1}
\end{figure}

\begin{algorithm}
\centering{ \fbox{
\begin{minipage}[t]{150mm}
\footnotesize
\setcounter{linecounter}{000}
\begin{tabbing}
aaaaaa \= \kill

\line{} \>  $v.exp ::  |\{u \in N(v) : u.state=In \}|$ \\

\line{} \> {\bf R1:}  $v.state=Out \wedge \forall u \in N(v) : (u.state=Out \wedge u.exp=0)   \longrightarrow v.state=In$  \\

\line{} \> {\bf R2:}  $v.state=In \wedge \exists u \in  N(v) : (u.state=In \vee u.exp>1) \longrightarrow v.state=Out$ \\

\end{tabbing}
\normalsize
\end{minipage}
} \caption{Maximal Distance-2 Independent Dominating Set - MD2IS}
\label{alg1}}
\end{algorithm}

\subsection{Closure}

\begin{lemma}
\label{lem1}
 When all the nodes are disabled, the set $S=\{ v \in V, v.state=In\}$ is maximal distance-2 independent set.

\end{lemma}

\begin{proofL}
Suppose that the system is in a legitimate configuration. Since
$R2$ is not enabled for nodes $v$ in $S$, the condition $(\exists
u \in  N(v) : (u.state=In \vee u.exp>1)$ is false. Thus, $\forall
u \in  N(v) : (u.state=Out \wedge u.exp \leq 1$). Therefore, at
distance 1 from $v$, all the nodes are out of $S$. And, at
distance-2 from $v$ there is no node in $S$ because $u.exp \leq 1$
which is exactly $=1$ ($u$ has at least a neighbor $v$ in $S$,
Therefore, all the neighbors of $u$ are out $S$ except $v$. This
implies that all the nodes in the neighborhood of $v$ at
distance-2 are out of $S$).

To show that $S$ is maximal, observe that if we want to add one
node $v$ with $v.state = Out$ then $v$ will have not all the
neighbors at distance-2 out of $S$ ($i.e.$ $v$ has at least a
neighbor at distance-2 in $S$). Thus, the addition of this node
will violate distance-2 independence.

\renewcommand{\toto}{lem1}
\end{proofL}

\subsection{Convergence}

\begin{lemma}
\label{lem2} If a node $v$ executes $R1$ becoming independent, it remains independent, and every node in its neighborhood at distance-2 still out of $S$ and cannot be enabled.
\end{lemma}

\begin{proofL}
When a node $v$ executes $R1$, that means all the nodes $u$ at distance-2 from $v$ are out of $S$. It is clear that no one of $u$ could enabled $R1$ because there exists at least $v$ in $S$ as a node in the neighborhood of $u$ at distance-2 from $u$.  Thus, $v$ will still in $S$ and all the neighbors at distance-2 still out of $S$.
\renewcommand{\toto}{lem2}
\end{proofL}

\begin{lemma}
\label{lem3}
Any node of $V$ will be enabled at most twice by R2 then R1. Thus, MD2IS terminates in the worst case at $2n$ moves under the expression model using a central daemon.
\end{lemma}
\begin{proofL}
Since Lemma \ref{lem2} shows that every node executes R1, it cannot move again. It follows that any node could be enabled (in the worst case) by only R2 and then R1. Consequently, $2n$ moves is an upper bound to stabilize for $n$ nodes.
\renewcommand{\toto}{lem3}
\end{proofL}

\begin{theorem}
\label{th1}
 MD2IS is a self-stabilizing algorithm that constructs Maximal Distance-2 Independent Set in $O(n)$ moves under expression model using a central daemon.
\end{theorem}

\begin{proofT}
The proof follows from Lemma \ref{lem1} and  Lemma \ref{lem3}.
\renewcommand{\toto}{th1}
\end{proofT}
The last theorem shows that MD2IS stabilizes under the central daemon and expression model. Now, we use the transformer proposed by \cite{Tur12} that gives a corresponding self-stabilizing algorithm MD2IS$^{D}$ which operates under distributed daemon and distance-one model.

\begin{theorem}
\label{th2}
MD2IS$^{D}$ converges into legitimate configuration in $O(nm)$ moves in the distance-one model under a distributed daemon. Where $m$ is the number of edges.
\end{theorem}

\begin{proofT}
Using Theorem \ref{th1}, the proof follows from Theorem 18 of \cite{Tur12}, where $m$ is the number of edges.
\renewcommand{\toto}{th2}

\end{proofT}

\section{Simulation results}
In this section, simulation tests are carried out in order to evaluate MD2IS. We calculate the cardinality of the MD2IS with MIS, and then, we observe how the cardinality of MD2IS is related when the graph size grows. Although the comparison with MIS is unfair, there is no other algorithm that we could refer to evaluate MD2IS cardinality. The algorithms are written in Java using expression model for MD2IS. For MIS \cite{GHJS03}, we use the implementation of Lukasz Kuszner \cite{Kus15} in which we have generated arbitrary graphs with different density having sizes from 500 nodes to 20000 nodes. For each size of graphs, we have carried out 5 to 10 executions and then we have taken the average value.

\begin{figure}[hbt!]
\begin{center}
\includegraphics[height=3in, width=5in]{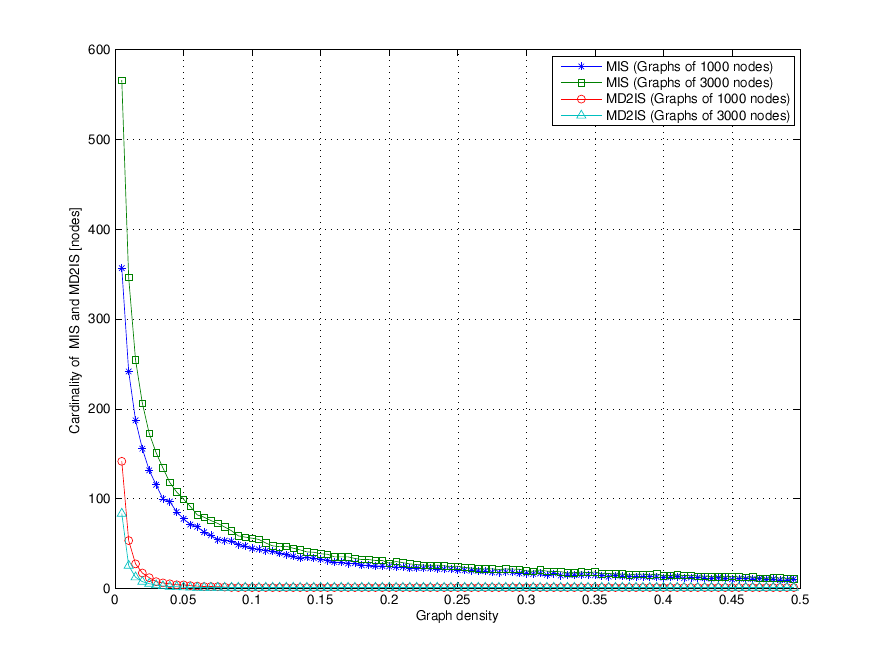}
\caption{Cardinality of MD2IS and MIS according density using graphs of 1000 and 3000 nodes.}
\end{center}
\label{fig1}
\end{figure}

Figure 3 shows the cardinality of MD2IS and MIS according graphs density. It is clear that MD2IS gives independent sets smaller than those produced by MIS. The density of the graphs has a clear impact on the cardinality of the independent sets. More the density grows, more the cardinality converges to be smaller. The important observation is that the cardinality of MD2IS will be close to 1 when the density becomes greater than 0.5. This is a rational result because for complete graphs (density=1), cardinality of MD2IS is 1.

\begin{figure}[hbt!]
\begin{center}
\includegraphics[height=3in, width=5in]{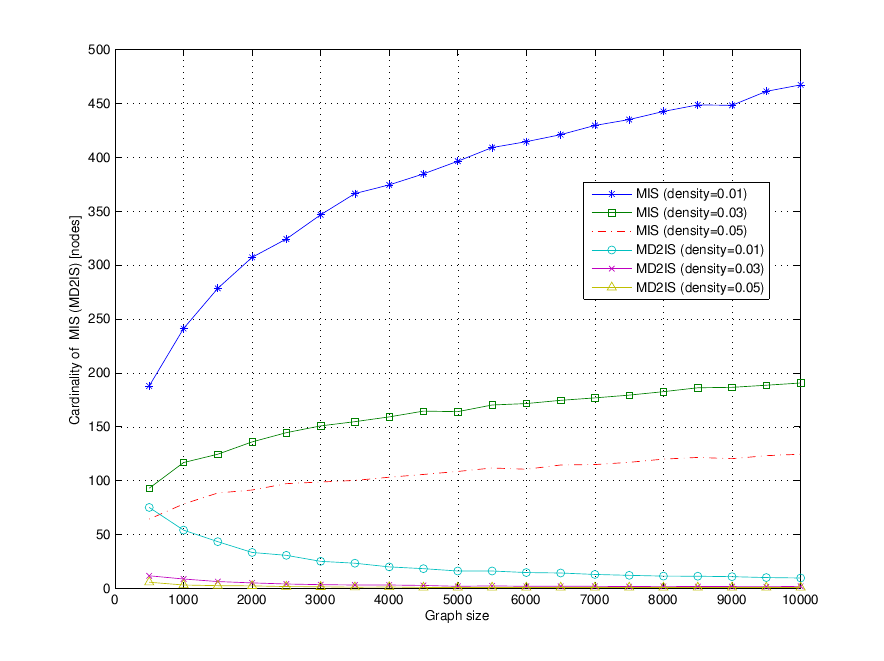}
\end{center}
\caption{Cardinality of MD2IS and MIS according graphs size
(Density=0.01).} \label{fig2}
\end{figure}
Figure 4 illustrates the cardinality of independent sets according size of graphs. Using constant density 0.01, curves show that MIS increases proportionally with the graph sizes. However, cardinality of MD2IS is inversely proportional to the graph sizes. For graphs of 10000 nodes, the cardinality of MIS is greater than 100 whereas the cardinality of MD2IS is less than 10 nodes.

\begin{table}[ht]
\begin{center}
\renewcommand{\baselinestretch}{1}
\small \noindent \begin{tabular}{|l|c|c|c|c|} \hline Graph size
 & {\bf MD2IS}  Cardinality  &{\bf MIS} Cardinality
&{\bf MD2IS} \noindent Convergence & {\bf MIS} Convergence \\

\hline \hline
 \textbf{1000}  &   601.8 (60.18\%)  &   687.8 (68.78\%)     &   400.2   &   415.6   \\
\hline
\textbf{1500}  &   703.6 (46.91\%)  &   926.4 (61.76\%)     &   652.8   &   636.6   \\
\hline
\textbf{2000}  &   758.2 (37.91\%)  &   1099.4 (54.97\%)    &   910.8   &   845.4   \\
\hline
\textbf{2500}  &   794.4 (31.78\%)  &   1248.2 (49.93\%)    &   1229.2  &   1078.4  \\
\hline
\textbf{3000}  &   786.6 (26.22\%)  &   1385.0 (46.17\%)    &   1570.8  &   1327.4  \\
\hline
\textbf{3500}  &   776.4 (22.18\%)  &   1522.4 (43.50\%)    &   1877.2  &   1624.6  \\
\hline
\textbf{4000}  &   762.2 (19.06\%)  &   1614.4 (40.36\%)    &   2202.8  &   1887.4  \\
\hline
\textbf{4500}  &   748.8 (16.64\%)  &   1720.4 (38.23\%)    &   2483.6  &   2206.0  \\
\hline
\textbf{5000}  &    724.2 (14.48\%) &   1801.0 (36.02\%)    &   2785.8  &   2454.8  \\
\hline
\textbf{5500}  &    711.0 (12.93\%) &   1887.0 (34.31\%)    &   3048.4  &   2749.2  \\
\hline
\textbf{6000}  &    692.2 (11.54\%) &   1964.4 (32.74\%)    &   3335.6  &   3061.6\\
\hline
\textbf{6500}  &    662.8 (10.20\%) &   2020.8 (31.09\%)    &   3551.6  &   3324.6\\
\hline
\textbf{7000}  &    642.6 (9.18\%)  &   2111.8 (30.17\%)    &   3825.2  &   3691,4\\
\hline
\textbf{7500}  &    633.8 (8.45\%)  &   2149.8 (28.66\%)    &   4098.8  &   3927.2\\
\hline
\textbf{8000}  &    604.6 (7.56\%)  &   2216.6 (27.71\%)    &   4333.4  &   4229.2\\
\hline
\textbf{8500}  &    601.8 (7.08\%)  &   2269.2 (26.70\%)    &   4586.4  &   4542.8\\
\hline
\textbf{9000}  &    577.6 (6.42\%)  &   2320.4 (25.78\%)    &   4836.0  &   4830.8\\
\hline
\textbf{9500}  &    552.8 (5.82\%)  &   2364.6 (24.89\%)    &   5097.2  &   5152.4\\
\hline
\textbf{10000} &    548.4 (5.48\%)  &   2415.2 (24.15\%)    &   5350.8  &   5379.2\\
\hline
\textbf{10500} &    527.4 (5.02\%)  &   2476.2 (23.58\%)    &   5645.0  &   5728.4  \\
\hline
\textbf{11000} &    501.6 (4.56\%)  &   2507.4 (22.79\%)    &   5817.6  &   6063.2\\
\hline
\textbf{11500} &    498.2 (4.33\%)  &   2560.4 (22.26\%)    &   6091.0  &   6303.6\\
\hline
\textbf{12000} &    481.2 (4.01\%)  &   2602.2 (21.69\%)    &   6300.6  &   6622.0\\
\hline
\textbf{12500} &    477.8 (3.82\%)  &   2635.2 (21.08\%)    &   6555.8  &   6923.8\\
\hline
\textbf{13000} &   454.6 (3.50\%)   &   2665.6 (20.50\%)    &   6790.0  &   7195.0\\
\hline
\textbf{13500} &   448.0 (3.32\%)   &   2705.0 (20.04\%)    &   7057.6  &   7496.6\\
\hline
\textbf{14000} &   435.2 (3.11\%)   &   2747.8 (19.63\%)    &   7251.6  &   7758.6\\
\hline
\textbf{14500} &   427.4 (2.95\%)   &   2785.0 (19.21\%)    &   7512.0  &   8045.0\\
\hline
\textbf{15000} &   416.2 (2.77\%)   &   2793.0 (18.62\%)    &   7798.4  &   8234.0\\
\hline
\textbf{15500} &   408.2 (2.63\%)   &   2839.4 (18.32\%)    &   7988.8  &   8579.8  \\
\hline
\textbf{16000} &   398.8 (2.49\%)   &   2867.4 (17.92\%)    &   8293.6  &   8857.8\\
\hline
\textbf{16500} &   392.0 (2.38\%)   &   2886.6 (17.49\%)    &   8496.2  &   9177.6\\
\hline
\textbf{17000} &   383.4 (2.26\%)   &   2934.2 (17.26\%)    &   8798.4  &   9433.6\\
\hline
\textbf{17500} &   380.2 (2.17\%)   &   2962.6 (16.93\%)    &   9043.4  &   9704.0\\
\hline
\textbf{18000} &   367.6 (2.04\%)   &   2976.6 (16.54\%)    &   9249.4  &   9972.0\\
\hline
\textbf{18500} &   361.0 (1.95\%)   &   3017.8 (16.31\%)    &   9481.0  &   10284.0\\
\hline
\textbf{19000} &   355.2 (1.87\%)   &   3036.6 (15.98\%)    &   9748.6  &   10535.8\\
\hline
\textbf{19500} &   353.0 (1.81\%)   &   3054.8 (15.67\%)    &   9998.8  &   10785.0\\
\hline
\textbf{20000} &   340.4 (1.70\%)   &   3080.2 (15.40\%)    &   10207.8 &   11032.4\\
\hline
\end{tabular}
\end{center}
\caption{Cardinality and time of convergence of MD2IS (on graphs
of density=0.001).} \label{table}
\end{table}

Further results of simulation are given in table 1 where the time of convergence is shown to be finite and variates proportionally to the graph size. Results confirm the lemma \ref{lem3} that algorithm MD2IS converges at most in $2n$ moves although the simulations gives smaller number of moves that is close to $n/2$.
In these tests, random graphs have been generated for small density = 0.001. We use this later value of density in order to be more close to real networks. For example, for a graph of 10000 nodes, a density of 0.001 will give an average degree = 10 for each node which models a user having 10 friends on social networks. The generated graphs have orders from 1000 nodes to 20000 nodes. We take the average value after carrying out 5 tests.

\section{Conclusion}

In this paper we proposed a first self-stabilizing MD2IS algorithm that converges into the correct configuration in $O(n)$ moves using a central daemon under expression model. In the distance-one model, MD2IS$^{D}$ terminates in $O(nm)$ moves using a distributed daemon.

The computation of the convergence time for independent sets of $k$ distance which still an open question is left for the future work. We plan also to evaluate the use of MD2IS as observers for the problem of locating sources propagating rumors on real graphs of social networks.

\clearpage

\bibliographystyle{abbrvurl}
\bibliography{mybibfile}



\end{document}